\documentclass[jou]{apa6}

\usepackage[natbibapa]{apacite}
\usepackage{graphicx}
\usepackage{mathtools}

\title{Power contours: optimising sample size and precision in experimental psychology and human neuroscience}

\author{Daniel H. Baker$^{1,2,6}$, Greta Vilidaite$^{3}$, Freya A. Lygo$^{1}$, Anika K. Smith$^{1}$, Tessa R. Flack$^{4}$, Andr\'e D. Gouws$^{5}$ \& Timothy J. Andrews$^{1}$}
        
\affiliation{1. Department of Psychology, University of York, Heslington, York, YO10 5DD, UK \linebreak
2. York Biomedical Research Institute, University of York, Heslington, York, YO10 5DD, UK \linebreak
3. School of Psychology, University of Southampton, University Road, Southampton, SO17 1BJ, UK \linebreak
4. School of Psychology, University of Lincoln, Brayford Pool, Lincoln, LN6 7TS, UK \linebreak
5. York Neuroimaging Centre, The Biocentre, York Science Park, Heslington, York, YO10 5NY, UK \linebreak
6. Corresponding author, email: \emph{daniel.baker@york.ac.uk}
}

\shorttitle{Power contours (preprint)}
\leftheader{Baker et al. (preprint)}
              
\abstract{When designing experimental studies with human participants, experimenters must decide how many trials each participant will complete, as well as how many participants to test. Most discussion of statistical power (the ability of a study design to detect an effect) has focussed on sample size, and assumed sufficient trials. Here we explore the influence of both factors on statistical power, represented as a two-dimensional plot on which iso-power contours can be visualised. We demonstrate the conditions under which the number of trials is particularly important, i.e. when the within-participant variance is large relative to the between-participants variance. We then derive power contour plots using existing data sets for eight experimental paradigms and methodologies (including reaction times, sensory thresholds, fMRI, MEG, and EEG), and provide example code to calculate estimates of the within- and between-participant variance for each method. In all cases, the within-participant variance was larger than the between-participants variance, meaning that the number of trials has a meaningful influence on statistical power in commonly used paradigms. An online tool is provided (https://shiny.york.ac.uk/powercontours/) for generating power contours, from which the optimal combination of trials and participants can be calculated when designing future studies.}


\begin{document}

\maketitle

\section{Introduction}
\label{intro}

\begin{figure*}
\begin{center} 
\includegraphics[width=0.95\textwidth]{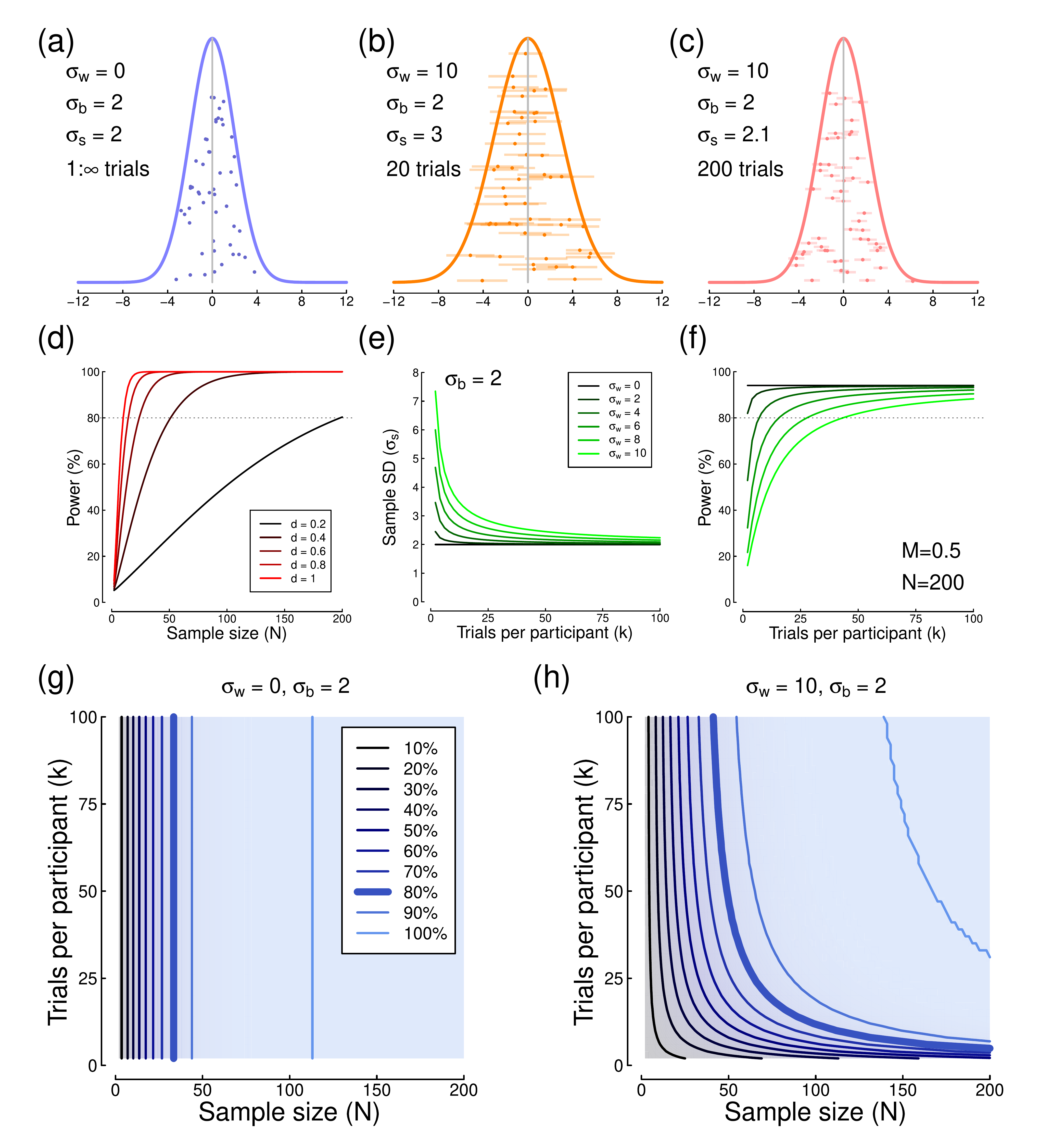}
\caption{Simulations of standard deviation and statistical power. Panel (a) shows simulated data for 50 individuals, generated using a population mean of $M = 0$, a within-participants standard deviation of $\sigma_{w}=0$, a between-participants standard deviation of $\sigma_{b}=2$, and a sample standard deviation of $\sigma_{s}=2$. Individual data points have a random vertical offset for display purposes. In panel (b) the within-participant standard deviation was increased to $\sigma_{w}=10$, and each point is the mean of 20 trials, with horizontal error bars indicating $\pm1$ SEM. Panel (c) shows the effect of increasing to 200 trials per participant. Panel (d) plots traditional power curves for different effect sizes (Cohen's \emph{d}) as a function of sample size (N). The dashed horizontal line indicates a power of 80\%, which is generally considered acceptable. Panel (e) shows how the sample standard deviation ($\sigma_{s}$) depends on the number of trials per participant (\emph{k}) for a range of within-participant standard deviations (see legend), and a between-participant standard deviation of $\sigma_{b}=2$. Panel (f) shows the statistical power resulting from the values in panel (e), for a sample size of $N=200$ and an underlying mean of $M = 0.5$. Panels (g,h) show power contours for different combinations of $\sigma_{w}$ and $\sigma_{b}$, as described in the text, and a group mean of $M=1$. Simulations used normally distributed random numbers, and statistical power was calculated for a two-sided, one-sample t-test comparing to a mean of 0.}
\label{fig1} 
\end{center}
\end{figure*}

Statistical power is the ability of a study design with a given sample size to detect an effect of a particular magnitude. In recent years, the problems with low statistical power have been increasingly highlighted \citep{Bishop2019}. Low powered studies are less able to detect a true effect (and so make more Type II errors) compared with high powered studies. Nominally significant findings from low powered studies are less likely to reflect true effects \citep{Button2013}, and because of publication bias (whereby significant findings are more likely to be published than non-significant ones) published low powered studies will also have a high Type 1 error (false positive) rate. Furthermore, any real effects that are detected are likely to have inflated effect sizes \citep{Ioannidis2008,Colquhoun2014}. These problems are common across many scientific disciplines, and estimates of power across studies in the neurosciences \citep{Button2013} yield power values in the range 8\%-30\%, far below the desired level of $\ge80\%$. The prevalence of low-powered studies has filled some areas of the literature with effects that fail to replicate and may well be spurious \citep{Ioannidis2005,OSC2015}. Most discussion of increasing statistical power has focussed on recruiting larger sample sizes, because for a given effect size, power is a function of sample size (see Figure \ref{fig1}d). However there is a second degree of freedom available to many experimenters at the study design stage -- the number of repetitions (or trials) of a given experimental condition by each participant. 

When the dependent variable of interest can be estimated with high precision, repeated measurements provide little benefit, and the main source of variance is between participants. This is illustrated by the distribution in Figure \ref{fig1}a, where participants (points) differ according to a normal distribution (curve), but the variance of each individual point is negligible. A more realistic situation for many experimental paradigms is shown in Figure \ref{fig1}b, where the variance of each individual estimate is large, as indicated by the horizontal standard error bars. This has the knock-on effect of increasing the overall standard deviation of the sample ($\sigma_{s}=2$ units in Figure \ref{fig1}a, and $\sigma_{s}=3$ units in Figure \ref{fig1}b). Such inflation of the sample standard deviation can be ameliorated by improving the accuracy of each participant's estimated mean by increasing the number of measurements. This is demonstrated in Figure \ref{fig1}c, where each participant's mean is estimated from $k=200$ trials (compared with $k=20$ in Figure \ref{fig1}b), and the standard deviation of the sample (curve) reduces substantially (to $\sigma_{s}=2.1$ units).

Power is typically derived using effect size measures such as Cohen's \emph{d} \citep{Cohen1988}, which depends on the sample mean (or difference in means), and also the sample standard deviation (formally $d = M/\sigma_{s}$). Under parametric assumptions, the number of trials per participant (\emph{k}) influences the sample standard deviation (Figure \ref{fig1}e), according to the equation:

\begin{equation}
\sigma_{s} = \sqrt{\sigma_{b}^2 + \frac{\sigma_{w}^2}{k}} 
\label{eq1}
\end{equation}

\noindent where $\sigma_{b}$ and $\sigma_{w}$ are the between- and (average) within-participant standard deviations, and \emph{k} is the number of trials per participant \citep[see also][]{Brandmaier2018}. The sample standard deviation ($\sigma_{s}$) determines the effect size, and subsequently the power (Figure \ref{fig1}f). In domains where the dependent variable is subject to high within-participant variance (as is potentially the case in psychology and neuroscience studies), increasing the precision of the per-participant estimate can therefore greatly increase overall power, perhaps reducing the number of participants required for a study \citep[see][]{Cleary1969,Phillips2016}. Although most active researchers are intuitively aware of this fact (it is common knowledge that running lots of trials delivers `better' data), and the problem has received mathematical treatment \citep{Williams1989,Kanyongo2007,Westfall2014,Phillips2016,Rouder2018}, there is no widely used procedure for quantitatively determining the appropriate number of trials to run. Instead, studies are typically designed using rules of thumb, prior precedent and guesswork.

In this paper, we advocate a useful representation, the \emph{power contour} plot -- a two-dimensional representation of power as the joint function of sample size ($N$) and number of trials ($k$). We provide an online tool for generating power contours in order to estimate the impact of measurement precision (the number of trials conducted) on statistical power. We then use a subsampling method to explore the joint effects of sample size and number of trials on real data sets using common methodologies and paradigms in psychology and neuroscience research. These measures include reaction times, psychophysical thresholds, event-related potentials, steady-state evoked potentials, and fMRI BOLD signals. We make computer code available to demonstrate how power contours were produced, and how estimates of the within- and between-participant variance were calculated for each example.

\section{Power contours}

Consider first the situation described above, in which the dependent variable of interest can be estimated accurately from a single trial, but individuals all express different true values of the variable (formally, the within-participant variance is low, but the between-participant variance is high, $\sigma_{w}<<\sigma_{b}$). Examples might include variables such as age and height, for which there is low measurement error and minimal variation from moment to moment, or for which tools exist (such as tape measures) to facilitate accurate measurement. In these situations, statistical power is a function of sample size and effect size (Figure \ref{fig1}d), where effect size is Cohen's \emph{d}. Clearly, in such a situation, testing each participant multiple times should confer no advantage. We can represent the power as a function of both sample size and number of trials using a 2D plot such as the one shown in Figure \ref{fig1}g. Here the lines trace iso-power contours - combinations of sample size and number of trials that result in the same statistical power \cite[this property is sometimes referred to as power equivalence, see][]{Oertzen2010}. For this example the power contours are vertical, showing no benefit of repeated testing.

Next consider a more realistic scenario -- a situation where the individual measurements are very noisy (high within-participant variance relative to the between-participant variance, $\sigma_{w}>\sigma_{b}$). The sample standard deviation decreases as a function of the number of trials (Figure \ref{fig1}e), as the estimated mean for each participant becomes more accurate with repeated measurements. Now power depends on both the number of trials and the sample size, and a series of curved iso-power contours are apparent (Figure \ref{fig1}h; see recent work by  \cite{Westfall2014} and \cite{Xu2018} for related plots in different scenarios).

These power contours offer a useful summary of the effect of possible experimental designs on statistical power. A given power (e.g. 80\%, indicated by the thick blue curves on the power contour plots) can be obtained from multiple combinations of sample size and trial number. This is a useful insight, as study designs can then be optimised depending on other constraints. If relatively few participants are available (perhaps because of financial constraints, or testing of a clinical population) then the number of trials can be increased. Note, however, that beyond a particular number of trials (around $k=50$ in Figure \ref{fig1}h), the function asymptotes and further trials are not beneficial. Alternatively, if each participant must be tested very rapidly (e.g. for studies involving children), but many participants are available, the number of trials could be kept relatively low (here around $k=20$), and a larger sample size tested. This is of potential value for large cohort studies, in which many participants each complete a large battery of various tasks. A more typical situation is one in which an experimenter wishes to minimise both sample size and testing time -- here values around the knee-point of the power contour permit joint optimisation of both parameters. Power contour plots can be produced for any combination of within- and between-participant variances and difference in means using an \emph{R} script, which can be accessed through a web interface at: https://shiny.york.ac.uk/powercontours/

To have practical value in the design of experiments, it is necessary to establish empirically whether power does indeed vary with the number of trials in typically used experimental paradigms. To this end, we have reanalysed data from 8 studies, using a range of common methodologies from psychology and cognitive neuroscience, including reaction times, proportional choices, sensory thresholds, EEG, MEG and fMRI. We estimate power contours by subsampling the data, so we aimed to include data sets featuring large sample sizes, in which each participant completed many trials (though it was not always possible to satisfy both criteria). All of these analyses are based on one-sample or paired t-tests, but the same principle applies to more sophisticated statistical techniques (see the Discussion section), and can be implemented using the subsampling technique we describe below. All analysis scripts are available on the project repository at https://osf.io/ebhnk/ and data sets are provided either on the project page or referenced directly throughout the manuscript to allow others to reproduce our analyses, and apply the methods to their own studies. We anticipate that these resources will be most valuable as a guide for performing related subsampling analyses for specific study designs, and suggest that readers short on time might find it most useful to skip ahead to the section reporting data from whichever paradigm they are most familiar.

\section{Reaction times}

\begin{figure*}
\begin{center} 
\includegraphics[width=0.95\textwidth]{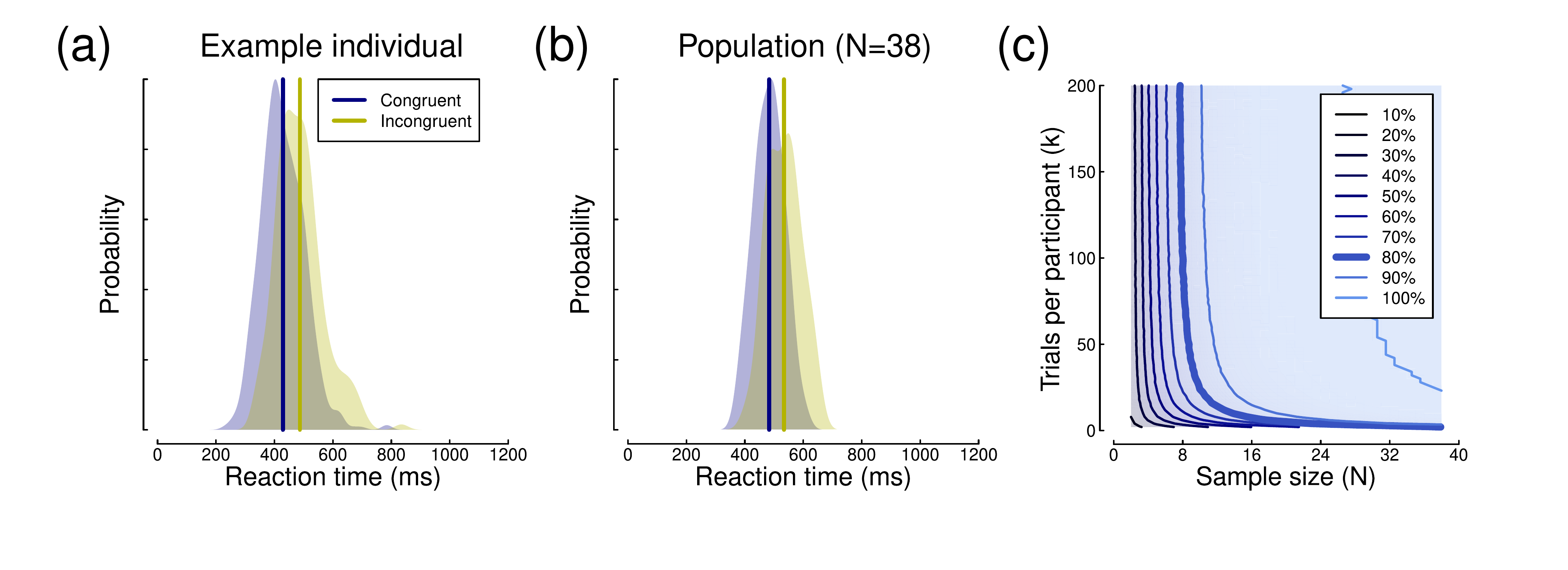}
\caption{Summary of reaction time data. Panel (a) shows reaction time distributions for an example participant, with vertical lines giving the means. Panel (b) shows the group level data for mean reaction times across the sample of 38 participants. Panel (c) shows a power contour plot, in which colour represents statistical power (see legend). The thick blue line indicates combinations of sample size and trial number with a power of 80\%. The y-axis represents the number of trials in the incongruent condition (the congruent condition contained three times as many trials)}
\label{fig2} 
\end{center}
\end{figure*}

We first analysed reaction time measures from a Posner-style attentional cueing experiment previously reported by \cite{Pirrone2018}. Participants (N = 38) saw a central cue stimulus directing their attention to either the left or the right of fixation. A sine wave grating target was then presented either in the attended location (congruent condition) or the unattended location (incongruent condition). Each participant completed $k=600$ congruent trials and $k=200$ incongruent trials, with example RT distributions for one participant shown in Figure \ref{fig2}a. At the group level, reaction times were on average 51 \emph{ms} slower in the incongruent condition (see Figure \ref{fig2}b), and the standard deviation of the differences ($\sigma_{s}$) was 42 \emph{ms}. For the full data set, this yielded an effect size of $d=1.2$. We also estimated the within participants standard deviation by pooling the variances for the incongruent and congruent reaction times, and averaging across participants, for which $\sigma_{w}$ = 151 \emph{ms}. Finally, to estimate $\sigma_{b}$ we rearranged equation \ref{eq1} to give:

\begin{equation}
\sigma_{b} = \sqrt{\sigma_{s}^2 - \frac{\sigma_{w}^2}{k}},
\label{eq2}
\end{equation}

\noindent which produced a value of $\sigma_{b}$ = 41 \emph{ms}.

We calculated statistical power by resampling random subsets of trials and participants from the data, and calculating the effect size and power using the mean and standard deviation, for a paired t-test comparing to 0 (using the \emph{pwr.t.test} function in the \emph{pwr} package in \emph{R}). Note that an alternative is simply to calculate a t-test with the resampled data, and calculate the proportion of tests that are significant, but the direct estimation of power is computationally more efficient so we use this where possible. The subsampling procedure was repeated 10,000 times, and the averaged power estimates are shown in Figure \ref{fig2}c. Just as predicted by our simulations (Figure \ref{fig1}h), the iso-power contour for 80\% power (shown by the thick blue line) is curved (we confirmed the subsampling result by using the summary statistics calculated above in the power contour \emph{Shiny} app). High power can be obtained with either a large sample size ($N>20$) and small number of trials ($k<10$) or a large number of trials ($k>50$) and small sample size ($N=8$). The knee-point of the function is around a sample size of $N=10$, with each participant completing approximately $k=20$ trials. Of course, this is for a relatively large effect size with a robust and well-established effect (attentional cueing). Other study designs with smaller sized effects will require larger sample sizes and/or more trials, but it is clear that the same basic pattern should apply for experiments of this type.

\section{Proportional choices in the Iowa Gambling Task}

\begin{figure*}
\begin{center} 
\includegraphics[width=0.95\textwidth]{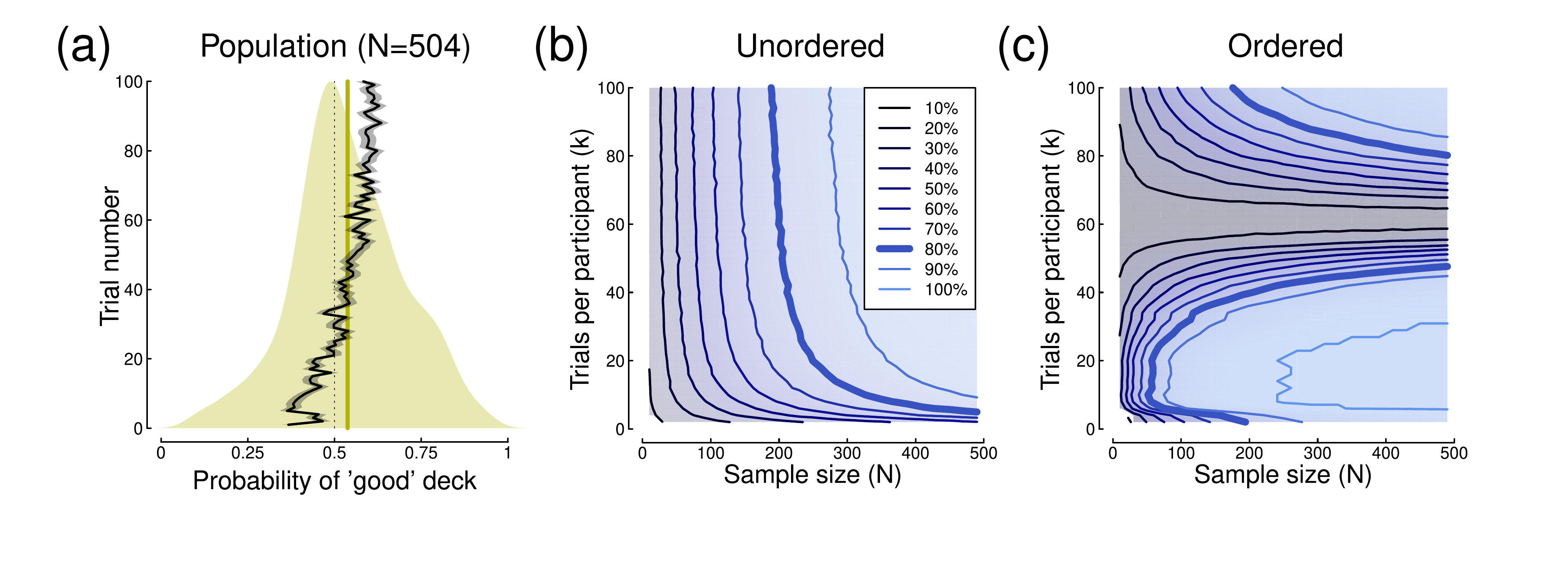}
\caption{Summary of proportion data from the Iowa Gambling task. Panel (a) shows a density plot of the mean probability of choosing a card from a `good' deck for the population of $N=504$ participants, each averaged across $k=100$ trials. The vertical orange line shows the grand mean, and the dashed vertical line is the probability expected by chance. The black curve (with grey shading showing $\pm$1SE) shows the mean probability across all participants on each trial (1 to 100). Panel (b) shows power contours for one-sample t-tests comparing the mean probability to the chance baseline (0.5). For these simulations, trials were randomly subsampled. Panel (c) shows power contours when trials were included sequentially.}
\label{fig3} 
\end{center}
\end{figure*}

We next reanalysed a data set comprising $N=504$ participants in the Iowa Gambling Task, as reported by \cite{Steingroever2015}, and made available through that publication. In this task, participants choose cards from four decks. Two decks have a greater overall payoff (`good' decks), and the other two have a poorer payoff (`bad' decks). Participants must learn these contingencies during the course of the experiment, and attempt to maximise their payoff. As such performance changes throughout the experiment, and we discuss the consequences of this learning below, but begin with an analysis of the aggregated (e.g. unordered) trials. Figure \ref{fig3}a shows summary data for a population of participants who each completed $k=100$ trials of the task. Averaged across all trials, the mean probability of selecting a card from a `good' deck was 0.54 (sample SD of $\sigma_{s}=0.16$), an effect size of $d=0.24$ when compared with the chance baseline of 0.5 (see Figure \ref{fig3}a). We calculated the standard deviation of individual choices, and averaged across participants to give $\sigma_{w}=0.47$, implying (via equation \ref{eq2}) a between-subjects standard deviation of $\sigma_{b}=0.15$.

We again calculated power by resampling random subsets of trials and participants from the data, and calculating the effect size and power using the mean and standard deviation, for a one-sample t-test comparing to 0.5 (using the \emph{pwr.t.test} function in the \emph{pwr} package in \emph{R}). This procedure was repeated 10,000 times, and the averaged power estimates are shown in Figure \ref{fig3}b. Consistent with the simulations in Figure \ref{fig1}h, power depends on both sample size and number of trials. With small numbers of trials ($k<40$), sample size can be dramatically reduced by increasing trial numbers. For example, by increasing from $k=5$ to $k=40$ trials, the sample size can be reduced from $N=400$ to $N=200$ whilst maintaining power. Alternatively, for a sample size of $N=200$, there are few gains to be made by increasing from $k=40$ to $k=100$ trials, as the function has reached asymptote.

In the Iowa Gambling Task, the trial contingencies are learned throughout the experiment. The black trace in Figure \ref{fig3}a illustrates that at the start of the experiment participants are more likely to choose cards from the `bad' decks for around the first 20 trials. Their behaviour then changes as they learn the task contingencies, and for the final 40 trials they are more likely to choose cards from the `good' decks. This information is lost by randomly sampling trials as we did to generate the power contour plot in Figure \ref{fig3}b. An alternative is to retain the trial order, and resample only across participants. Power contours are shown for this analysis in Figure \ref{fig3}c. Over the first 40 trials, power is high because the mean probability is significantly below 0.5 (see black curve in Figure \ref{fig3}a). As participants start to learn the task contingencies, the mean probability passes through 0.5, and power falls to near zero around 60 trials. Then, as participants begin to reliably choose the `good' deck, the average probability becomes significantly above 0.5 and power increases again, reaching 80\% by around 80 trials with the full sample of participants. This alternative visualisation of the data could be valuable when planning studies using this task, as it  shows explicitly how performance (and hence overall power) changes over time.

\section{Sensory thresholds}

\begin{figure*}
\begin{center} 
\includegraphics[width=0.95\textwidth]{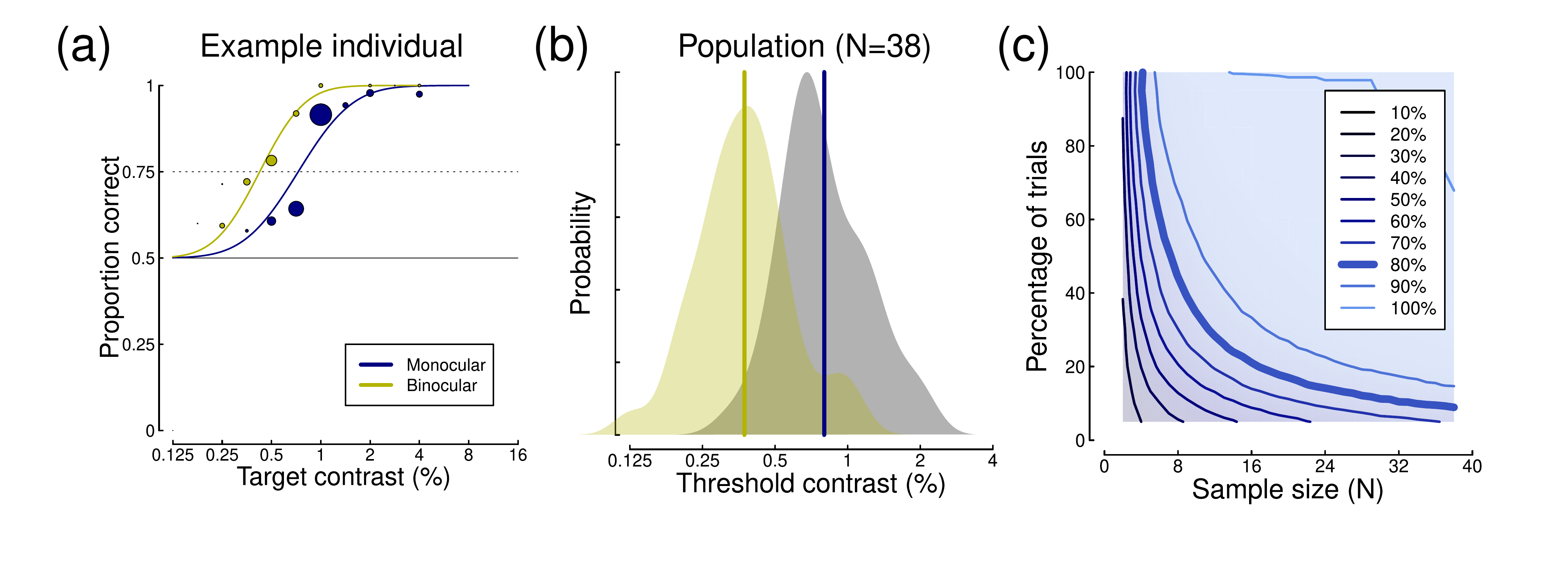}
\caption{Summary of threshold psychophysics data. Panel (a) shows psychometric functions for a single participant, with symbol size proportional to the number of trials at each target contrast level. Curves are fitted cumulative Gaussian functions, used to interpolate thresholds at 75\% correct (dashed line). Data for the monocular condition (blue) were pooled across the left and right eye conditions before fitting. Panel (b) shows distributions of monocular (blue) and binocular (yellow) detection thresholds across a group of $N=38$ participants with normal vision. Panel (c) shows the power contours derived by subsampling the data and refitting the psychometric functions.}
\label{fig4} 
\end{center}
\end{figure*}

Psychophysical detection thresholds are typically measured using large numbers of binary trials across stimuli of different intensities. The proportion of correct trials increases monotonically with stimulus intensity, producing a psychometric function (see Figure \ref{fig4}a). Threshold is then estimated at some criterion performance level (often 75\% correct) by fitting a continuous ogival function such as a cumulative Gaussian or Weibull distribution. We reanalysed data from a binocular summation experiment \citep[reported by][]{Baker2018}, in which contrast detection thresholds were measured in this way for sine wave grating stimuli shown either monocularly or binocularly using a stereo shutter goggle system. Example psychometric functions for a single participant are shown in Figure \ref{fig4}a  \citep[fitted using the \emph{quickpsy} package in \emph{R}, see][]{Linares2016}, where it is clear that equivalent performance requires higher contrast for monocular presentation (blue) than for binocular presentation (yellow). At the group level (see Figure \ref{fig4}b), this produces a ratio of monocular to binocular thresholds between $\sqrt{2}$ and 2 -- the well-known binocular summation effect -- which here had an effect size of $d=1.8$. The mean effect was $6.6dB$, with a sample standard deviation of $\sigma_{s}=3.6dB$

We subsampled the data set to produce the power contour plot shown in Figure \ref{fig4}c. Because each participant completed slightly different numbers of trials (owing to the adaptive staircase procedure used to determine contrast levels for each trial), we subsampled at different percentages of trials for each participant, refitting the psychometric function each time. On average, each participant completed 225 trials for the binocular condition, and for the monocular conditions for each eye (left and right eyes were tested separately and their data combined). Summation estimates were rejected when they fell outside of a reasonable range (between factors of 0.12 and 32), as this indicated that something had gone wrong with the fitting procedure. As anticipated, power depended on both sample size and number of trials, and continued to improve over the ranges available in the data set (i.e. the function at 80\% power was quite shallow, and did not asymptote over the ranges tested). Indeed, with all trials included, only around six participants were required to reach 80\% power \citep[consistent with previous estimates of power for this paradigm, see][]{Baker2018}. Conversely, when all 38 participants were included, only around 15\% of trials were required (around 34 trials for each condition). Alternatively, 80\% power could be maintained with a sample size of $N=12$, with each participant completing around 30\% of the total trials.

For this paradigm, estimating the within-participant standard deviation was not straightforward because threshold were calculated by fitting a psychometric function. So, we generated power contour surfaces for a range of possible $\sigma_{w}$ values, and compared these numerically to the surface derived by subsampling (Figure \ref{fig4}c). The best fitting value was $\sigma_{w}=33.5dB$, which implies (via equation \ref{eq2}) a between-participant standard deviation of $\sigma_{b}=1.3dB$.

\section{EEG: event-related potentials}

\begin{figure*}
\begin{center} 
\includegraphics[width=0.95\textwidth]{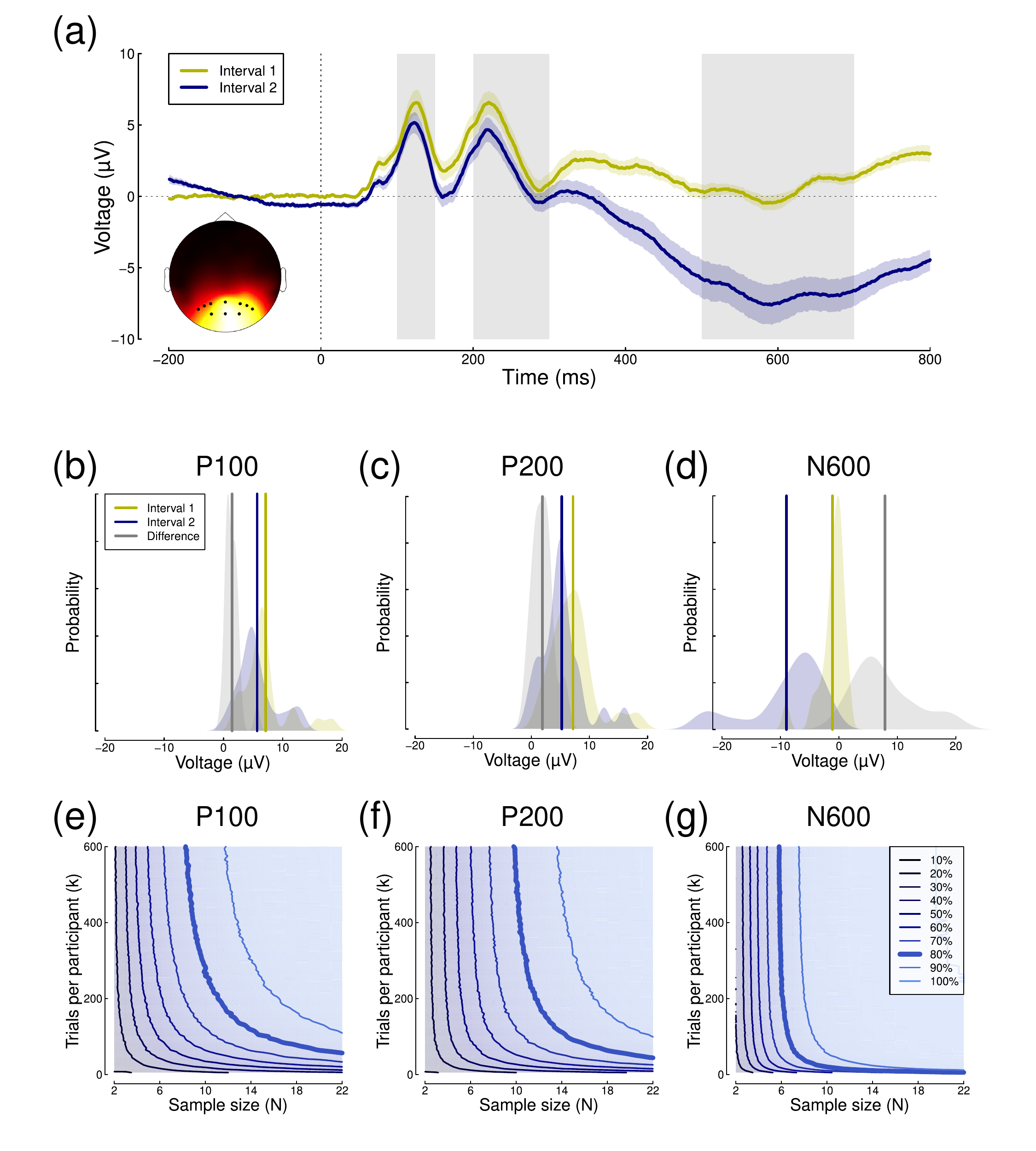}
\caption{Summary of ERP results. Panel (a) shows grand mean ERPs in response to central presentation of a 50\% contrast sine wave grating in two intervals of each trial. Shaded regions surrounding each trace show $\pm$1SE across participants ($N=22$), and the grey rectangles illustrate the time windows used to estimate peaks. The inset shows the distribution of voltages across the scalp at 226 \emph{ms} after stimulus onset and black symbol mark the electrodes (\emph{Oz, O1, O2, POz, PO3 - PO8}) over which ERPs were averaged. Panels (b-d) show average peak voltages across a group of $N=22$ participants in each time window, for both intervals and their difference. Panels (e-g) show power contours for the peak voltage within each time window.}
\label{fig5} 
\end{center}
\end{figure*}

We next analysed event-related potentials (ERPs) from a contrast discrimination experiment reported by \cite{VilidaiteERP2019}, recorded using a 64-channel EEG cap. The stimuli were sine wave gratings with a contrast of 50\%, presented sequentially in pairs for 100 \emph{ms} each, with an interstimulus interval of 400-600ms. These produced a typical response (see Figure 5a) over occipital electrodes (see inset to Figure \ref{fig5}a), with positive peaks at around 120 and 220 \emph{ms} (marking stimulus onset and offset), and a later negative region with a trough around 600 \emph{ms}. The first stimulus of each pair (yellow curve) produced a generally more positive response than the second stimulus (blue curve), in part as a consequence of differential overlap, though the precise cause of the differences are unimportant for this demonstration. Each trial was baselined by subtracting the mean voltage during the 200 \emph{ms} before stimulus onset. The sample size for this experiment ($N=22$) was modest (albeit typical for ERP research), but each participant completed a large number of trials ($k=600$ stimulus pairs). 

For each participant, we calculated the peak voltage and latency within three time windows, highighted grey in Figure \ref{fig5}a. These were 100-150 \emph{ms}, 200-300 \emph{ms} and 500-700 \emph{ms}, and corresponded to the P100, P200 and N600 components. The peak voltages and latencies were compared between the two intervals using a repeated measures approach. The distributions of peak voltages and voltage differences across participants are shown in Figure \ref{fig5}b-d for the three time windows, which produced effect sizes (Cohen's \emph{d}) of 1.18, 1.11 and 1.32. We performed similar calculations for the latencies, however these were less convincing, with effect sizes of \emph{d}=0.21, 0.04 and 0.47 for the three time windows. We do not consider them further here, though power contours could be calculated for data sets with more robust latency differences.

We calculated power contours for each of the three peak voltage differences by subsampling trials and participants, and re-estimating the peak for each participant and condition on each of 10,000 iterations. These are shown in Figure \ref{fig5}e-g, and had the expected format in all cases. For the P100 component, power continued to increase across all sample sizes and trial numbers tested. For the N600 component, power was largely determined by sample size, and only for relatively few trials ($k<200$) could sample size be materially reduced by adding more trials. This suggests that the limitations on statistical power in typical ERP experiments can depend on both sample size and number of trials, and that their relative contributions may depend on the size of the effect being studied. See also \cite{Boudewyn2018} and \cite{Clayson2017} for more detailed discussion of these issues in ERP studies. For this data set, estimates of standard deviations ranged from $12\mu V$ to $21\mu V$ for $\sigma_{w}$, and from $1.1\mu V$ to $5.3\mu V$ for $\sigma_{b}$.

\section{EEG: steady-state evoked potentials}

\begin{figure*}
\begin{center} 
\includegraphics[width=0.65\textwidth]{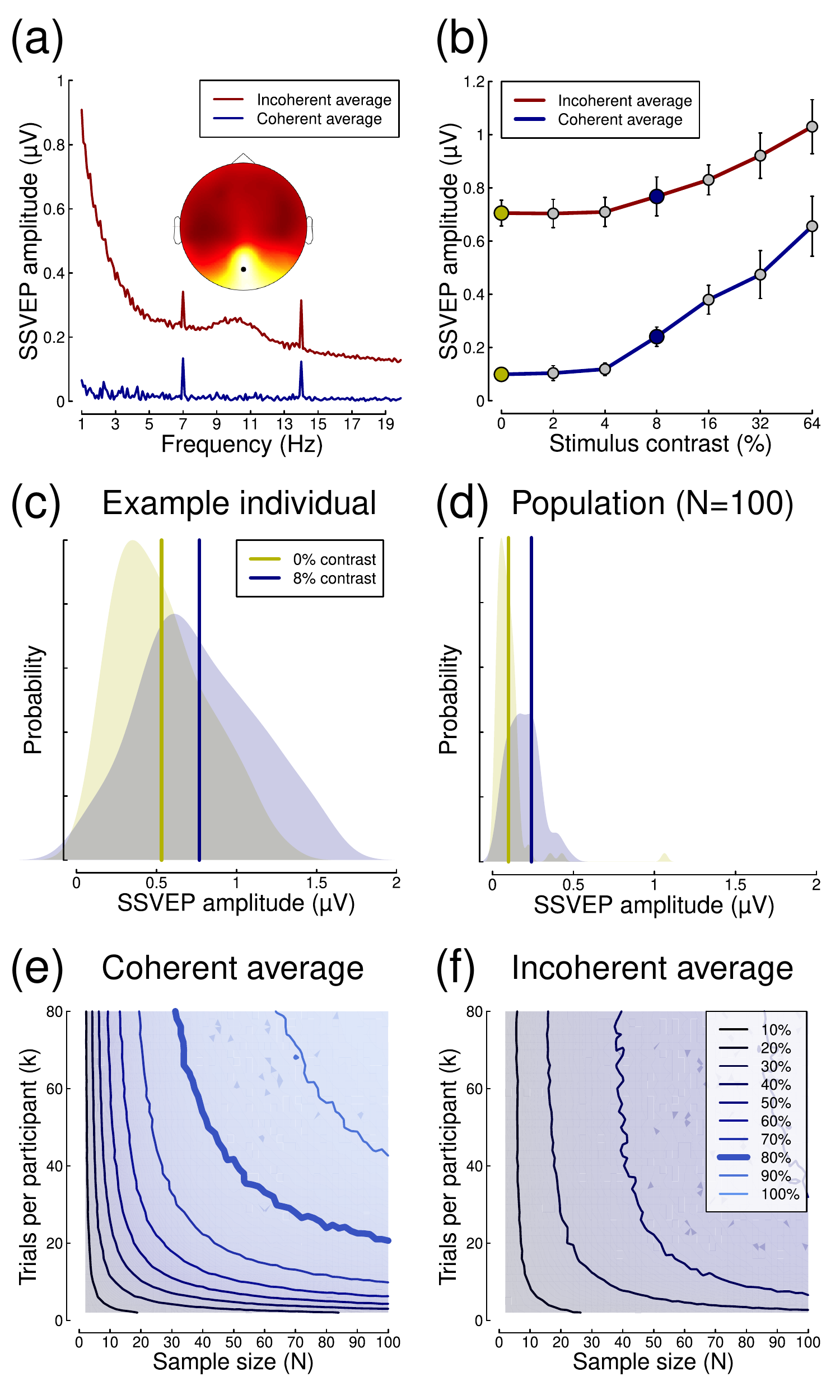}
\caption{Summary of SSVEP data. Panel (a) shows Fourier spectra for full 10 \emph{s} long trials, using either coherent (blue) or incoherent (red) averaging, and the scalp distribution of activity at 7Hz (inset). Panel (b) shows contrast response functions for both types of averaging. Panel (c) shows the distribution of amplitudes for an example participant, and panel (d) shows averages for the population. Panels (e) and (f) show power contours for coherent and incoherent averaging, respectively.}
\label{fig6} 
\end{center}
\end{figure*}

An alternative EEG paradigm is the steady-state method, where a stimulus oscillates at a particular frequency, inducing entrained neural responses at that same frequency. In an experiment reported by \cite{Vilidaite2018}, sine wave gratings of different contrasts were flickered at 7Hz, and shown to a sample of $N=100$ participants. Each participant completed 8 trials of 11 seconds per contrast level, from which the first $1s$ of EEG data was discarded, and the remaining $10s$ were divided into 10 epochs of $1s$ each, yielding a total of $k=80$ observations per condition. Each epoch was then Fourier transformed, and responses are evident both at the fundamental (flicker) frequency (7Hz) and its second harmonic (14Hz), as shown in Figure 6a. For these visual stimuli, the responses are strongest at the occipital pole, near early visual cortex (see inset to Figure \ref{fig6}a).

Responses at the fundamental frequency increase monotonically with maximum stimulus contrast (see Figure \ref{fig6}b) at electrode \emph{Oz}. For a stimulus contrast of 8\% (marked by the blue circle), comparing absolute responses (i.e. removing the phase component before averaging) to the baseline condition (0\% contrast, yellow circle) results in an effect size of $d=0.2$. However this can be substantially increased (to $d=0.68$) by using coherent averaging, in which both the amplitude and phase information are averaged across trials for each individual participant (and the absolute amplitudes are then averaged across participants). The improvement occurs because responses to the stimuli are phase-locked, and therefore should have the same phase on each trial. Any noise at the stimulus frequency has random phase, and so cancels out over multiple repetitions. Example Fourier spectra for both coherent (blue) and incoherent (red) averaging methods are shown in Figure \ref{fig6}a, where it is clear that the coherent method greatly reduces the noise at off-target frequencies. Note in particular the increase in noise in the alpha band (8-12Hz) is clear with incoherent averaging (red) but absent with coherent averaging (blue). In the contrast response function (Figure \ref{fig6}b), coherent averaging (blue function) leads to lower amplitudes at low stimulus contrasts, whereas with incoherent averaging (red function) responses must overcome a much higher `noise floor' before they can be detected. Distributions of voltages for an example participant and for the population are shown in Figure \ref{fig6}c,d.

We calculated power contours via subsampling using both coherent (Figure \ref{fig6}e) and incoherent (Figure \ref{fig6}f) averaging, which further confirmed that coherent averaging results in substantially greater statistical power. The 80\% power contour in the coherent condition (thick line in Figure \ref{fig6}e) is relatively shallow, showing that both increasing sample size and adding more trials will improve power over most of the range explored here. For example, halving the sample size from $N=100$ to $N=50$ requires an increase from approximately $k=20$ to $k=40$ trials per participant to maintain power at 80\%. We confirmed these general findings at the higher stimulus contrasts (not shown). Because the coherent averaging precludes typical calculation of within-participant standard deviations, we again fitted the power contour surfaces for a range of $\sigma_{w}$ to the power contours derived by subsampling. The best fitting values were $\sigma_{w}=3.1\mu V$ and $\sigma_{b}=0.19\mu V$.

\section{fMRI: event-related design}

A widely-used fMRI paradigm is the event-related design, in which stimuli are presented briefly with a jittered interstimulus interval (ISI). We obtained data from the Cam-CAN repository (available at http://www.mrc-cbu.cam.ac.uk/datasets/camcan/) for an event-related fMRI experiment detailed by \cite{Shafto2014} and \cite{Taylor2017}. In brief, $N=625$ participants viewed bilateral checkerboard patterns, presented for 30 \emph{ms} and repeated $k=124$ times. Some stimuli were accompanied by an auditory beep, but this was disregarded for the purposes of our analyses.

\begin{figure*}
\begin{center} 
\includegraphics[width=0.95\textwidth]{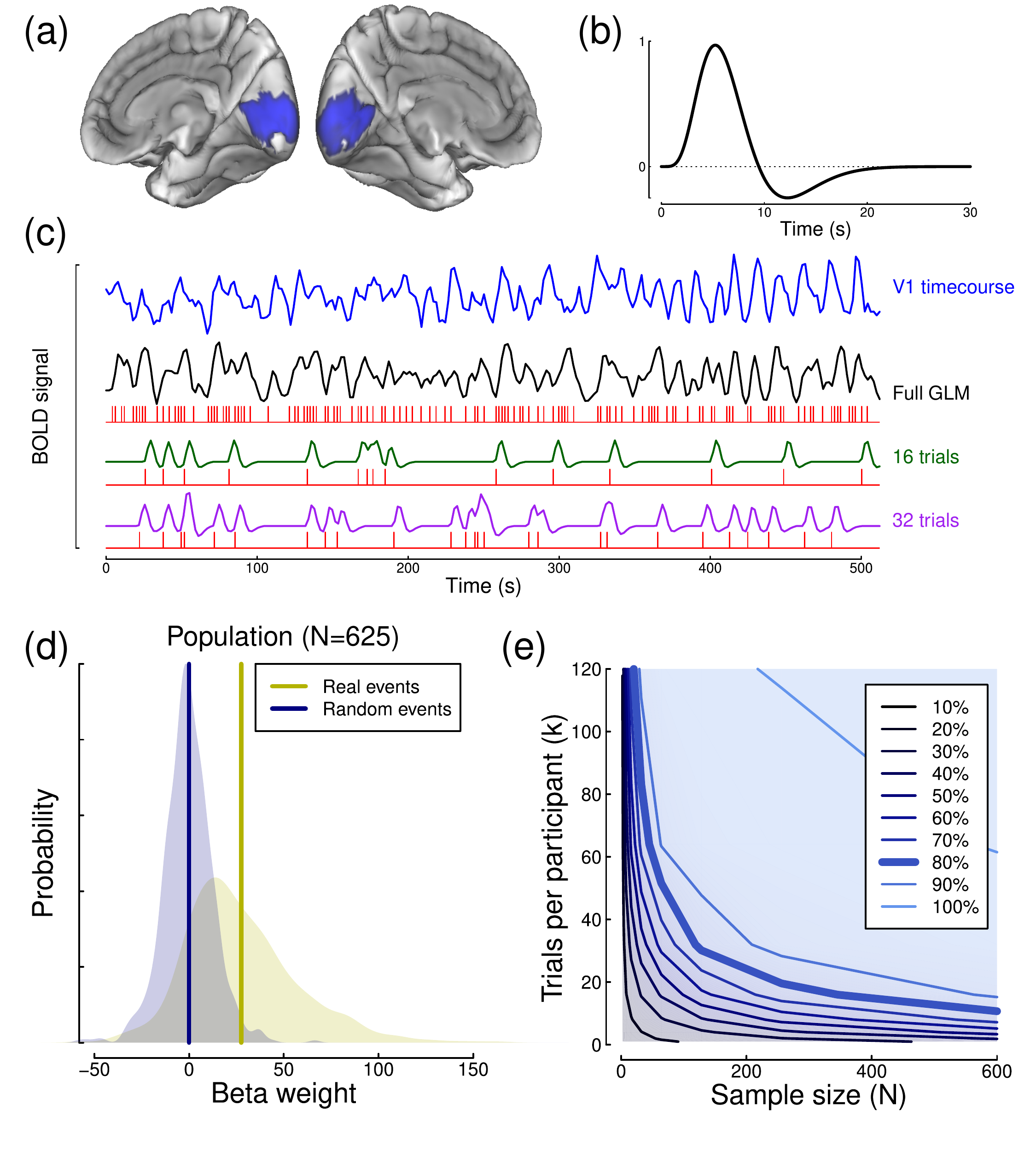}
\caption{Summary of event-related fMRI analysis and results. Panel (a) shows the V1 region of interest on the medial surface of the standard (MNI152) brain, highlighted in blue. Panel (b) shows the canonical double gamma haemodynamic response function used in our general linear models. Panel (c) shows an example time-course from the V1 ROI for one participant (blue), and a general linear model constructed to predict this time-course (black) based on stimulus events (red). The green and purple traces show example GLM components with random subsets of trials. Panel (d) shows the population distributions of beta weights for the full GLM modelling all stimulus events (yellow) or randomly simulated times (blue). Panel (e) shows the power contour plot for these event-related fMRI data.}
\label{fig7} 
\end{center}
\end{figure*}

We implemented a minimal preprocessing pipeline using FSL \citep{Jenkinson2012}. This involved co-registering the functional data to an individual participant's anatomical scan, and then to the standard MNI152 brain. We used the inverse of these transforms to project a probabilistic map of primary visual cortex (V1) obtained from \cite{Wang2015} onto the functional data to use as a region of interest (see Figure \ref{fig7}a). The functional data were corrected for slice timing and participant motion, and high pass filtered at 0.01Hz. Then the time-course was averaged across the V1 ROI and exported for further analysis. Whilst this anatomically-defined ROI will necessarily include some voxels that were not responsive to the stimulus, we would expect noise from these voxels to average out and not adversely affect the results \citep[e.g.][]{Boynton1996}.

We then constructed general linear models (GLMs) for each data set using the individual trial timings. To simulate experiments with variable numbers of trials, each GLM split the data using random trial allocations into two arbitrary groups -- a `target' condition and a 'non-target' condition. A third condition modelled four auditory-only trials which lacked any visual stimulus. A canonical double gamma haemodynamic response function (Figure \ref{fig7}b) was convolved with each condition using the \emph{fmri.stimulus} function \citep[part of the \emph{fmri} package in \emph{R}, see][]{Tabelow2011}, and orthogonal second order polynomial drift terms were included in the overall model. We then fit the GLM to determine a regression (beta) weight for the target condition to use as our dependent variable. By varying the number of trials allocated to the target and non-target conditions, we were able to simulate experiments with different numbers of trials, whilst keeping the GLM design balanced (see Figure \ref{fig7}c). To provide a null condition, we repeated the analysis using randomly determined events within the experiment time-course (i.e. not using the true event timings). This generated the sample distributions of beta weights shown in Figure \ref{fig7}d, and resulted in an effect size of $d=0.9$ for the full data set.

We calculated effect sizes across participants for the difference between beta values for the true and null models with different numbers of trials (see Figure \ref{fig7}c), and used these to estimate statistical power. As previously, simulations were repeated 10,000 times with different random sampling of trials and participants to generate power contours (see Figure \ref{fig7}e). As with several previous data sets, power continued to increase across the full range of trial numbers, such that 80\% power could be maintained for sample sizes from $N=20$ to $N=600$, simply by varying the number of trials. This flexibility allows event-related designs to achieve high statistical power even with relatively modest sample sizes, but it is critical that sufficient trials are included for each condition. It is also straightforward to design a severely underpowered study by including too few trials (here $k<60$). We estimated standard deviations by fitting to the subsampled power contour surface, yielding values of $\sigma_{w}=515$ and $\sigma_{b}=32.2$ (in $\beta$ units).

\section{fMRI: blocked design}

Another popular fMRI paradigm is the blocked design, in which stimuli are presented for periods of several seconds, interleaved with periods of no stimulation. Typically, events are scheduled to coincide with the acquisition of functional volumes (the repetition time, or TR). Blocked designs generally have greater power than event-related designs, because the stimulus timing is more closely aligned to the sluggish time constraints of haemodynamic activity, with the longer duration presentations (relative to event-related designs) allowing BOLD signals to sum over time \citep{Boynton1996}. 

We reanalysed a data set comprising N=83 participants, all of whom viewed a series of images of faces, objects, places and scrambled images as part of a functional localiser described by \cite{Flack2015}. Stimuli were presented in blocks of 6 \emph{s}, with a 9 \emph{s} inter-block interval during which the display was blank. Within each block, 5 images were shown sequentially for 1000 \emph{ms} each, with a 200 \emph{ms} inter-stimulus interval. fMRI data were acquired with a TR of 3 \emph{s}, so a complete cycle (one block plus inter-block interval) lasted for 15 \emph{s}, or 5 TRs. Each participant completed $k=35$ blocks. Functional data were high pass filtered, detrended and converted to percent signal change, and aligned to the MNI152 brain. The timeseries was then averaged across the V1 ROI shown in Figure \ref{fig7}a. 

\begin{figure*}
\begin{center} 
\includegraphics[width=0.95\textwidth]{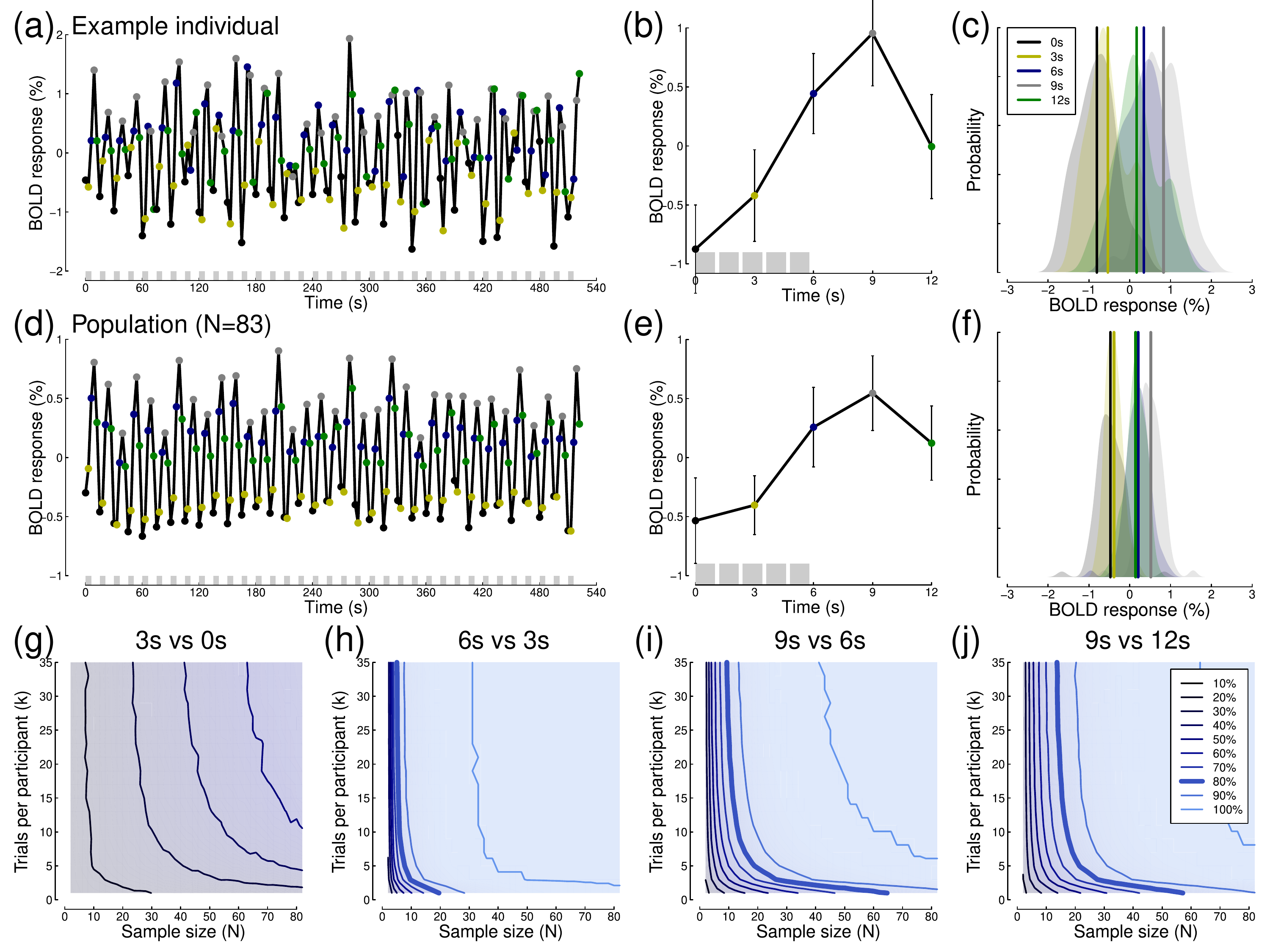}
\caption{Summary of blocked design fMRI data. Panel (a) shows an fMRI timecourse for an example individual, averaged across the V1 ROI (see Figure \ref{fig7}a). Shaded grey regions at the foot of the panel indicate blocks when stimuli were presented. Panel (b) shows the data from panel (a) aligned to each block onset and averaged across all $k=35$ blocks (with error bars showing $\pm$1SD). The grey shaded regions at the foot of the panel indicate the presentations of individual stimuli within a block. Panel (c) shows distributions of BOLD activity at each time point. Panels d-f mirror panels a-c but for the sample of $N=83$ participants. Panels g-j show power contours for the fMRI data, comparing activity at successive time points.}
\label{fig8} 
\end{center}
\end{figure*}

A timeseries for an example participant is shown in Figure \ref{fig8}a, and exhibits clear stimulus-driven modulations, with a period of 15 \emph{s} matching that of the trial cycle. The BOLD response peaked 9 \emph{s} after stimulus onset, as can be seen most clearly in Figure \ref{fig8}b, which averages the response across all 35 blocks for the example participant. The distributions of BOLD responses at each time point (relative to the start of a block) are shown in Figure \ref{fig8}c. Panels d-f of Figure \ref{fig8} show comparable data for the population of $N=83$ participants, displaying a similar pattern. In order to generate power contours for a range of effect sizes, we compared activity between sequential pairs of sample points. Effect sizes increased from $d=0.26$ comparing 3 \emph{s} and 0 \emph{s}, to $d=1.7$ comparing 6 \emph{s} and 3 \emph{s}. The range of standard deviations across these comparisons for $\sigma_{w}$ was 0.47 - 0.52\%, and for $\sigma_{b}$ was 0.23 - 0.40\%. Power contours (see Figure \ref{fig8}g-j) approximately asymptoted for trial numbers above $k=15$. This pattern is somewhat different from the event-related fMRI results discussed previously (Figure \ref{fig7}), where adding more trials continued to increase power across the entire range. For the larger effects (Figure \ref{fig8}h-j), power was high even with the relatively small samples ($N<20$) typical of many neuroimaging studies \citep{Button2013}. Of course looking for responses to visual stimuli in V1 is guaranteed to produce large effect sizes - most fMRI studies are designed to test subtler effects which will inevitably be smaller than in the examples here.

\section{MEG: evoked responses}

\begin{figure*}
\begin{center} 
\includegraphics[width=0.8\textwidth]{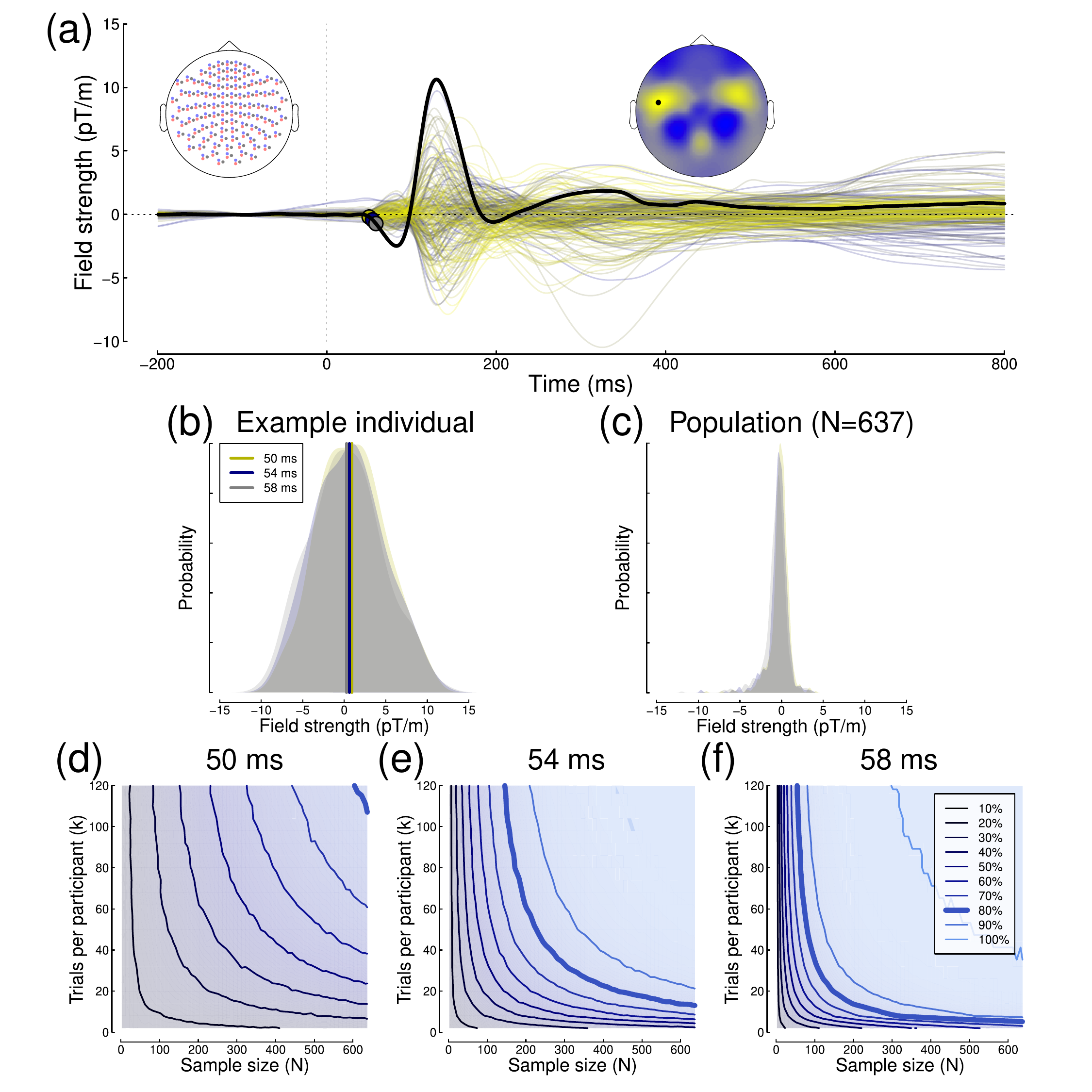}
\caption{Summary of MEG results. Panel (a) shows a butterfly plot of evoked responses from 204 planar gradiometers, averaged across all participants ($N=637$). The MEG montage is depicted in the upper left inset, where planar gradiometers of orthogonal orientations are indicated in blue and red, and magnetometer locations are shown in grey. The upper right inset shows the distribution of field strengths across a subset of 102 gradiometers with consistent orientation at 130 \emph{ms} (the peak of the black curve), and the black dot indicates the location of the sensor used for the analysis. Coloured points highlighted on the black curve indicate time points used for power analysis. Panel (b) shows distributions of field strengths at each of the three target time points for an individual participant. Panel (c) shows the same but for the sample population of $N=637$ participants. Panels (d-f) show power contours for different time-points.}
\label{fig9} 
\end{center}
\end{figure*}

The Cam-CAN data set also contains MEG responses ($k=120$ trials) to the same visual stimuli as described in the section on event-related fMRI, recorded using a \emph{VectorView} system (Elekta Neuromag, Helsinki). We filtered (0.01 - 30Hz bandpass), baselined and epoched the data from each participant, and then conducted one-sample t-tests at a single sensor (see Figure \ref{fig9}a) comparing activity to zero. We selected three time points very soon after stimulus onset (50, 54 and 58 \emph{ms}) to leverage the power of this large ($N=637$) dataset, and to explore effects of a similar magnitude to those investigated in typical experiments, where small differences in responses to different stimuli or mental states might be compared.

Evoked responses showed an initial polarisation beginning around 50 \emph{ms}, followed by a larger peak of opposite polarity at 130 \emph{ms} (see Figure \ref{fig9}a). Effect sizes at the three time points increased from $d=0.17$ at 50 \emph{ms} to $d=0.51$ at 58 \emph{ms}  when including all trials and participants. As for previous examples, the within-participant variance (Figure \ref{fig9}b) was clearly greater than the sample variance (Figure \ref{fig9}c). Across the time window from $50 - 400ms$, values of $\sigma_{w}$ ranged from $8.25 - 11.77pT/m$, and values of $\sigma_{b}$ ranged from $0.87 - 6.61pT/m$. Subsampled power contours showed the familiar form (see Figure \ref{fig9}d-f), with power only reaching 80\% for the 50 \emph{ms} time-point when the full data set was used. At later time points, iso-power contours show constant power can be maintained, for example when reducing the sample size from N = 400 to N = 200 by increasing the number of trials from $k=20$ to $k=60$ (at 54 \emph{ms}).

\section{Discussion}

\begin{figure*}
\begin{center} 
\includegraphics[width=0.75\textwidth]{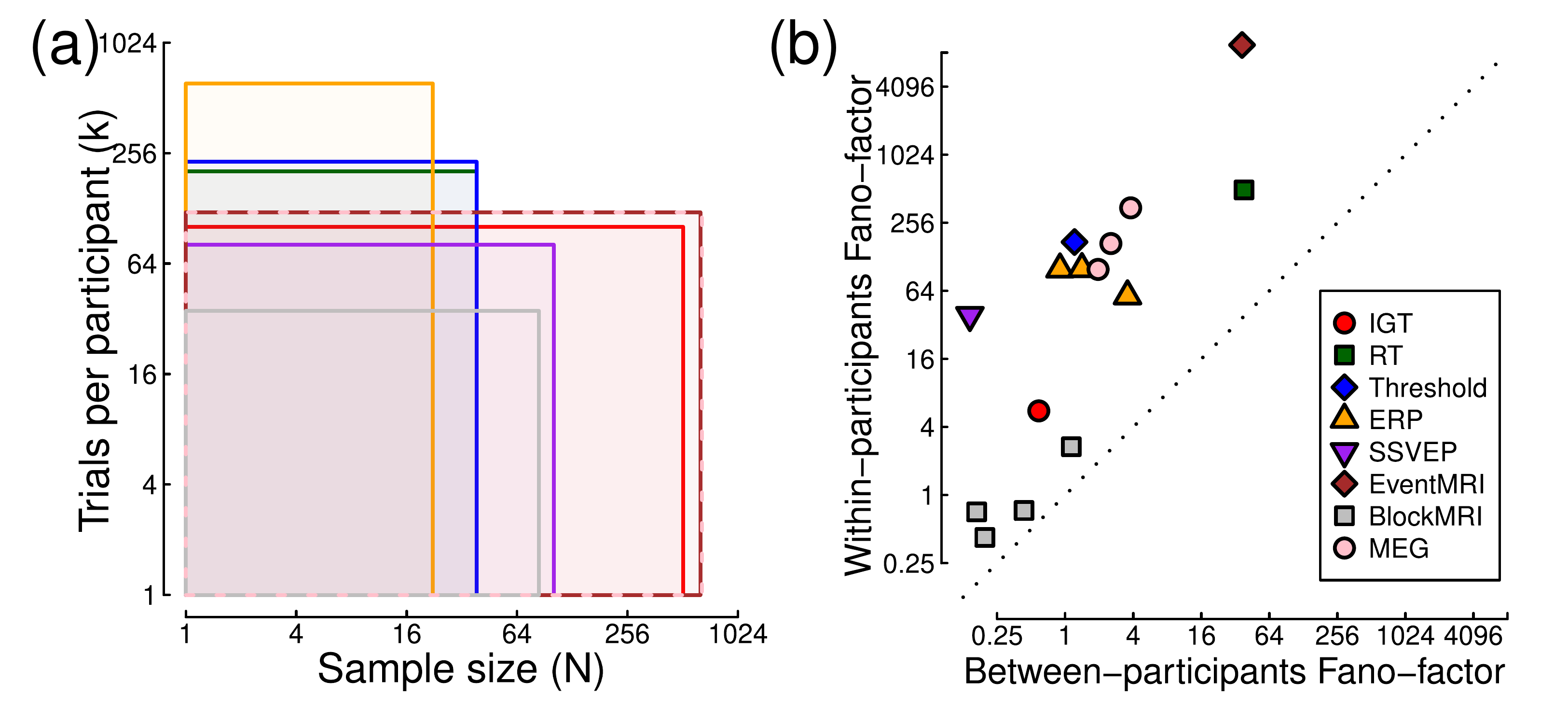}
\caption{Summary of sample sizes, trial numbers and Fano-factors across experimental paradigms. Each rectangle in (a) covers the range of sample sizes and trial numbers for one of the studies analysed here, with colours defined in the legend in panel (b). Panel (b) plots Fano-factors (variance divided by the mean) derived from the within- and between-participants standard deviations given in Table \ref{table:effects}. Note the log-scaled axes for both panels.}
\label{fig10} 
\end{center}
\end{figure*}

We advocate a representation of statistical power as the joint function of sample size and number of trials; the power contour plot. Example power contours were generated by subsampling data sets from a number of widely used paradigms in experimental psychology and human neuroscience, covering a range of different sample sizes and trial numbers (summarised in Figure \ref{fig10}a). In most cases, iso-power contours revealed situations where statistical power could be maintained with fewer participants, provided that each participant completed a larger number of trials. For some paradigms, power reached asymptote at a particular number of trials, beyond which further testing conferred no benefit for assessing statistical significance (though as we note below, additional trials may be informative in studies of individual differences). In other paradigms, particularly those where the dependent variable was derived by some form of model fit, power continued to improve with repeated testing, beyond the range that could be assessed with our data sets.

A practical guide to using the power contour approach for study design is as follows. If existing data are available on which to base an analysis, and where these data permit direct estimation of mean difference, $\sigma_w$ and $\sigma_b$ (using equation \ref{eq2}), these values can be calculated (or estimated using bootstrapping methods, see \cite{Luck2019}) and entered directly into the power contour web application. Where direct estimation of these values is not possible, power contours should be generated by subsampling, as we have done for the examples here (and as demonstrated in the code provided). If required, the effective values of $\sigma_w$ and $\sigma_b$ can then be estimated by fitting the subsampled power contour surface to simulated surfaces and finding the best fitting values. These methods will be of most use when planning replication studies, or when conducting a series of experiments using a single technique that build upon an initial finding in a well-powered sample. If no relevant data are available, power contours can still be informative if reasonable assumptions can be made about the likely effect size, and ratio of standard deviations. Just as it is common practise in power analysis to calculate power curves for a range of potential effect sizes, it might also prove instructive to compare power contour plots for a range of assumptions about the underlying effect size and variance measures. In all cases, the accuracy of the predictions will be limited by the extent to which the parameters generalise to the new experiment.

In Table \ref{table:effects}, we summarise the relevant variables from each paradigm, including the mean effect, and within- and between-participant and sample standard deviations. For several paradigms, including sensory thresholds, SSVEPs, and event-related fMRI, estimates of within-participant standard deviations were not directly available because the process by which trials were combined did not generate one. In these cases (as described above), we simulated power contour surfaces for a range of candidate standard deviations. The estimated value is the within-participant standard deviation ($\sigma_{w}$) that produced the best fit. Although this has no direct relationship to the measured dependent variable, it can be thought of as the SD from an experimental design with identical power \cite[a power equivalent model, see][]{Oertzen2010} but which uses traditional averaging across trials instead of more sophisticated analysis steps. We then calculated the between-participant standard deviation ($\sigma_{b}$) using equation \ref{eq2}. For the SSVEP and event-related fMRI data sets, equation \ref{eq2} returned an imaginary number because the estimated within-participant SD was very large. Here we assumed that $\sigma_b=\sigma_s$ for the purposes of completing Table \ref{table:effects}. The analysis scripts used to perform these calculations are available on the project OSF repository (https://osf.io/ebhnk/), and we anticipate that readers might use these resources to perform similar analyses on their own data when planning future studies. However we advise caution in the extent to which variance estimates can be assumed to generalise across different experimental set-ups, laboratories, and participant groups. Using the values estimated here to perform power analyses for studies using similar methods is likely to be highly inaccurate and we do not recommend it.

\begin{table*}[ht]
\caption{Summary of means, standard deviations and effect sizes for different paradigms. The SD ratio is defined as $\sigma_{w}/\sigma_{b}$. *Estimated: to estimate a within-participant SD, we ran simulations to optimise the value of this parameter using the full power contour surface, see text for details.}
\centering
\begin{tabular}{l | c c c c c c}
Paradigm & Mean effect & $\sigma_{w}$ & $\sigma_{b}$ & SD ratio & $\sigma_{s}$ & Effect size (\emph{d}) \\
\hline
Reaction times & 51 \emph{ms} & 151 \emph{ms} & 41 \emph{ms} & 3.6 & 42 \emph{ms} & 1.2 \\
Iowa Gambling Task & 0.04 & 0.47 & 0.15 & 3.1 &  0.16 & 0.24 \\
Sensory thresholds & 6.6 dB & 33.5 dB* & 1.3 dB & 11.9 & 3.6 dB & 1.8 \\
ERP P100 & 1.47 $\mu$V & 12.0 $\mu$V* & 1.14 $\mu$V & 10.5 & 1.25 $\mu$V & 1.2 \\
ERP P200 & 1.93 $\mu$V &  13.8 $\mu$V* & 1.64 $\mu$V & 8.4 & 1.74 $\mu$V & 1.1 \\
ERP N600 & 7.84 $\mu$V &  21.1 $\mu$V* & 5.27 $\mu$V & 4.0 & 5.34 $\mu$V & 1.5 \\
SSVEP 8\% vs 0\% & 0.25 $\mu$V & 3.1 $\mu$V* & 0.19 $\mu$V & 16.3 & 0.19 $\mu$V & 0.7 \\
Event-related fMRI & $\beta = 28.6$ & $\beta = 515$* & $\beta = 32.2$ & 16.0 & $\beta = 32.2$ & 0.9 \\
Blocked fMRI 3 \emph{s} vs 0 \emph{s} & 0.09\% & 0.49\% & 0.32\% & 1.5 & 0.33\% & 0.26 \\
Blocked fMRI 6 \emph{s} vs 3 \emph{s} & 0.59\% & 0.50\% & 0.34\% & 1.5 & 0.35\% & 1.70 \\
Blocked fMRI 9 \emph{s} vs 6 \emph{s} & 0.31\% & 0.47\% & 0.23\% & 2.1 & 0.24\% & 1.29 \\
Blocked fMRI 9 \emph{s} vs 12 \emph{s} & 0.37\% & 0.52\% & 0.40\% & 1.3 & 0.41\% & 0.91 \\
MEG 50 \emph{ms} & 0.20 \emph{p}T/\emph{m} & 8.25 \emph{p}T/\emph{m} & 0.87 \emph{p}T/\emph{m} & 9.5 & 1.15 \emph{p}T/\emph{m} & 0.17 \\
MEG 54 \emph{ms} & 0.42 \emph{p}T/\emph{m} & 8.32 \emph{p}T/\emph{m} & 1.03 \emph{p}T/\emph{m} & 8.1 & 1.28 \emph{p}T/\emph{m} & 0.32 \\
MEG 58 \emph{ms} & 0.72 \emph{p}T/\emph{m} & 8.38 \emph{p}T/\emph{m} & 1.18 \emph{p}T/\emph{m} & 7.1 & 1.41 \emph{p}T/\emph{m} & 0.51 \\

\end{tabular}
\label{table:effects}
\end{table*}

A further instructive analysis is to compare the within- and between-participant variances, as these provide insight into the likely gains that can be obtained by conducting more trials on each participant. A situation in which the within-participant variance is very small compared to the between-participants variance will result in a power contour like that shown in Figure \ref{fig1}g, where repeated testing confers no benefit. Figure \ref{fig10}b plots the variances expressed as Fano-factors (variance scaled by the mean) to permit comparison across paradigms with widely differing units. It is clear that for all paradigms considered here, the within-participant variance is substantially above the between-participant variance (all points appear above the diagonal). This property is not a given, and we anticipate that there may exist paradigms where within-participant variance is very low (owing to accurate measurement, or consistency of responses across multiple repetitions; see \cite{Nesselroade1991} for a discussion in the context of developmental research). We note that where multiple estimates were calculated for a single method (such as ERPs at different time points), the Fano-factors appear to cluster together, suggesting a consistent ratio of variances for a given paradigm. However, establishing a generic Fano factor for a particular methodology would require further investigation across multiple studies, and also across different laboratories and equipment (e.g. scanner models, sensor types etc), and would not necessarily apply to individual experiments.

From equation \ref{eq1}, the sample standard error can be expressed as:

\begin{equation}
{SE}_s = \sqrt{\frac{\sigma^2_b+\frac{\sigma^2_w}{k}}{N}} = \sqrt{\frac{\sigma^2_b}{N}+\frac{\sigma^2_w}{kN}}.
\label{eq3}
\end{equation}

\noindent These expressions make explicit the dependence of measurement precision (and hence power) on both \emph{N} and \emph{k}, regardless of effect size. In situations where $\sigma_{w}>\sigma_{b}$, running many trials will materially reduce the overall standard error. In situations where $\sigma_{w}<\sigma_{b}$, running many trials will confer less benefit, as the standard error is primarily determined by $\sigma_{b}$, and increasing $N$ is more profitable. In Table \ref{table:effects} we also calculate the ratio of standard deviations ($\sigma_{w}/\sigma_{b}$) as this gives a useful indication of the likely influence that changing $k$ will have on power. Paradigms with a small ratio (such as the blocked fMRI paradigm) produce power contours with the smallest gains from increasing numbers of trials (see Figure \ref{fig8}).

\begin{figure*}
\begin{center} 
\includegraphics[width=0.75\textwidth]{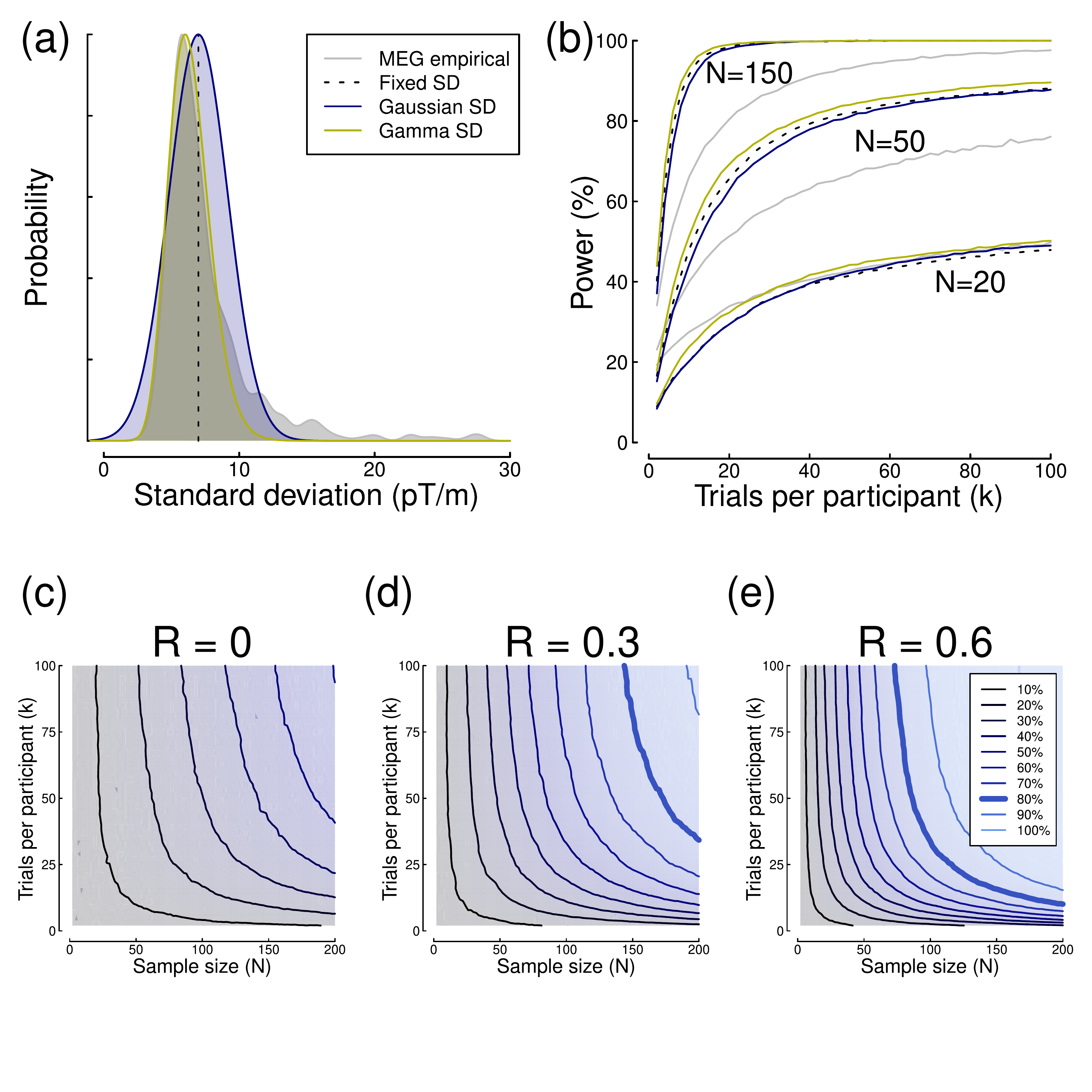}
\caption{Summary of the influence on power of the distribution of within-participant standard deviations, and the correlation between repeated measures. Panel (a) illustrates possible distributions of within-participant standard deviations. The grey curve shows an empirical distribution derived from the MEG data set (N = 637 at 58 \emph{ms}). The dashed line gives a fixed value, which is the mean of the empirical distribution excluding values >15 \emph{pT/m}. The blue curve shows a normal distribution, with mean and SD derived from the empirical distribution (mean = 6.99, SD = 2.17). The yellow curve shows the  gamma distribution that best fits the empirical distribution (shape = 17.64, scale = 0.36). Panel (b) shows statistical power as a function of the number of trials for a range of sample sizes, using the four distributions shown in (a). Panels (c-e) show simulated power contours for repeated measures designs as a function of the correlation (R) between the two conditions. For these simulations we assumed a group mean difference of 0.5, between participants standard deviation of 2, and within participant standard deviation of 10. The total variance remained constant across the range of correlations.}
\label{fig11} 
\end{center}
\end{figure*}

Up until this point, we have implicitly assumed that a fixed value of within-participant standard deviation ($\sigma_{w}$) can be substituted for each participant's individual value. Is this assumption justified, and what impact might different distributions of $\sigma_{w}$ have on statistical power? To address this, we simulated power curves assuming a fixed value of $\sigma_{w}$, and both normal and skewed distributions of $\sigma_{w}$ (see Figure \ref{fig11}a). The properties of these distributions were derived from the MEG data set (at 58 \emph{ms}), as described in the Figure \ref{fig11} caption, and compared with power estimates from the empirical data. For a range of sample sizes (N) and numbers of trials (\emph{k}), the power estimates for all three artificial distributions were very similar (Figure \ref{fig11}b). However the power estimates derived from the empirical data are somewhat lower, especially with larger numbers of participants. This happens because a small number of outlier participants with higher standard deviations (those in the tail of the grey distribution in Figure \ref{fig11}a) contribute disproportionately to the overall variance. We think that most analysis pipelines will reject such participants (or reject individual trials that are contributing to a noisy participant mean), meaning that the loss of power here is a 'worst case' scenario (we avoided elaborate processing pipelines in the current paper to maximise transparency). In general these simulations suggest that the simplifying assumption of a single within-participant standard deviation is reasonable. For prospective power analyses, the margin of error in estimating effect sizes and variances will most likely subsume any considerations owing to non-normally distributed variances and outliers.

A further factor that influences statistical power in repeated measures designs is the covariance between the two measures. We performed simulations to quantify this, by generating synthetic data sets with different levels of covariance, and performing power calculations on the synthetic data for repeated measures t-tests. The simulations in Figure \ref{fig11}c-e show that when R = 0, there is no benefit from the repeated measures design, and power is determined by conventional factors (effect size, alpha level, sample size and number of trials). As the level of correlation increases from zero, power also increases because the covariance between the two measures accounts for a greater proportion of the total variance, and it is discounted by the repeated measures analysis. However the overall shape of the power contours is not affected by the change in covariance - the contours simply shift towards the origin. For paired t-test designs, the covariance can be accounted for by taking the difference between the two measures for each participant, and using these difference scores in a one-sample t-test (which is mathematically equivalent to a paired t-test on the original data). Estimates of effect size and power calculated in this way will incorporate the covariance between repeated measures. For more sophisticated designs, calculating stochastic power contours using existing data, or simulating them with a range of plausible covariance levels, may be more appropriate.

Of course, we are far from the first to appreciate that multiple measurements can increase effect sizes and power. In the domain of psychometric research, the Spearman-Brown prophecy formula \citep{Spearman1910,Brown1910} predicts how the reliability of a test (such as a personality test, or an IQ test) increases as more items are added. \cite{Rouder2018} also consider the effects of sample size and number of trials on statistical power, in the context of `stochastic dominance' - the tendency for all participants in an experiment to have a true effect in the same direction. Under these conditions, the distribution of effects in the sample population is unlikely to be normal, and may instead be positively skewed with a mean and variance that are proportional (e.g. a gamma distribution). Simulations show that in this situation power can remain almost constant when trading off participants against trials. Our observation that Fano-factors for a given method appear to cluster together (see Figure \ref{fig10}b) could be taken as evidence that dominance holds for some of the paradigms investigated here, because gamma distributions have a variance that increases in proportion to the mean. Strong empirical evidence to firmly establish the conditions when dominance occurs is currently lacking, although it appears entirely plausible for many tasks in sensory and cognitive research.

Whereas most of the example data sets we consider here involve multiple repetitions of identical stimuli (6/8 used simple patterns such as checkerboards or sine-wave gratings), it is more typical in some research areas to use different stimulus examples on each trial. For example in research on object processing, databases of object images are often used, with multiple examplars in each object category. This additional source of variability can also be estimated, and further complicates the underlying mathematics of power analysis, as described in detail by \cite{Westfall2014}. The power contour representation advocated here is also applicable to these situations \citep[see Figures 2-6 of][]{Westfall2014}, and a linear mixed modelling approach can be used in which variances are explicitly represented at the participant, stimulus item and sample level \citep[see also][]{Brysbaert2018}. In such `crossed' designs, the maximum power that can be achieved is limited by the item-level variance and number of stimulus examples, even for a hypothetically infinite sample size. For statistical procedures where the item-level variance is not explicitly modelled, it will be subsumed into the within- and between-participant variances, perhaps making power estimates less accurate.

Some studies have used cost functions to attempt to derive a single optimal experimental design, by assuming specific costs (usually in units of experimenter time) required for recruitment and testing of each participant \citep[e.g.][]{Cleary1969,Oertzen2010,Oertzen2013}. In principle these methods might be used to determine a point on the power contour that specifies a particular sample size and number of trials. We have avoided being prescriptive about this here, as different studies will have different constraints and priorities, and the advantage of visualising the entire power surface is that it permits the experimenter to trade off these two variables against each other without loss of power. However we have built functionality into the \emph{Shiny} web application to estimate an optimal combination of sample size and number of trials, based on the additional constraint of a per-participant `recruitment cost', expressed as a notional number of trials. The optimal point is calculated by determining the smallest value of N*(\emph{k} + cost) that achieves 80\% power. We advise caution in the use of this feature.

\subsection{Application to other statistical tests and approaches}

Throughout all examples so far we have deliberately used a basic statistical test to determine power - the t-test. However the subsampling method we develop here can very easily be extended to more advanced statistical methods, including nonparametric statistics, Analysis of Variance \citep[see][for a related example]{Smith2018}, correlation, regression and so on. The method of subsampling trials has no specific requirements about the form of the data (as with bootstrapping techniques), provided the assumptions for calculating the relevant test statistic are met. A recent study by \cite{Xu2018} calculated the reliability of working memory measures as a function of both sample size and number of trials, using a similar sub-sampling approach. This produced similar contour plots, but for Cronbach's alpha, Spearman-Brown reliability and standard deviation instead of statistical power. In all cases, these showed a dependency on both sample size and number of trials, consistent with the examples here. Iso-power contours have also been calculated in work on optimal study design using structural equation modelling \cite[e.g.][]{Oertzen2013,Brandmaier2015}.

\begin{figure*}
\begin{center} 
\includegraphics[width=0.75\textwidth]{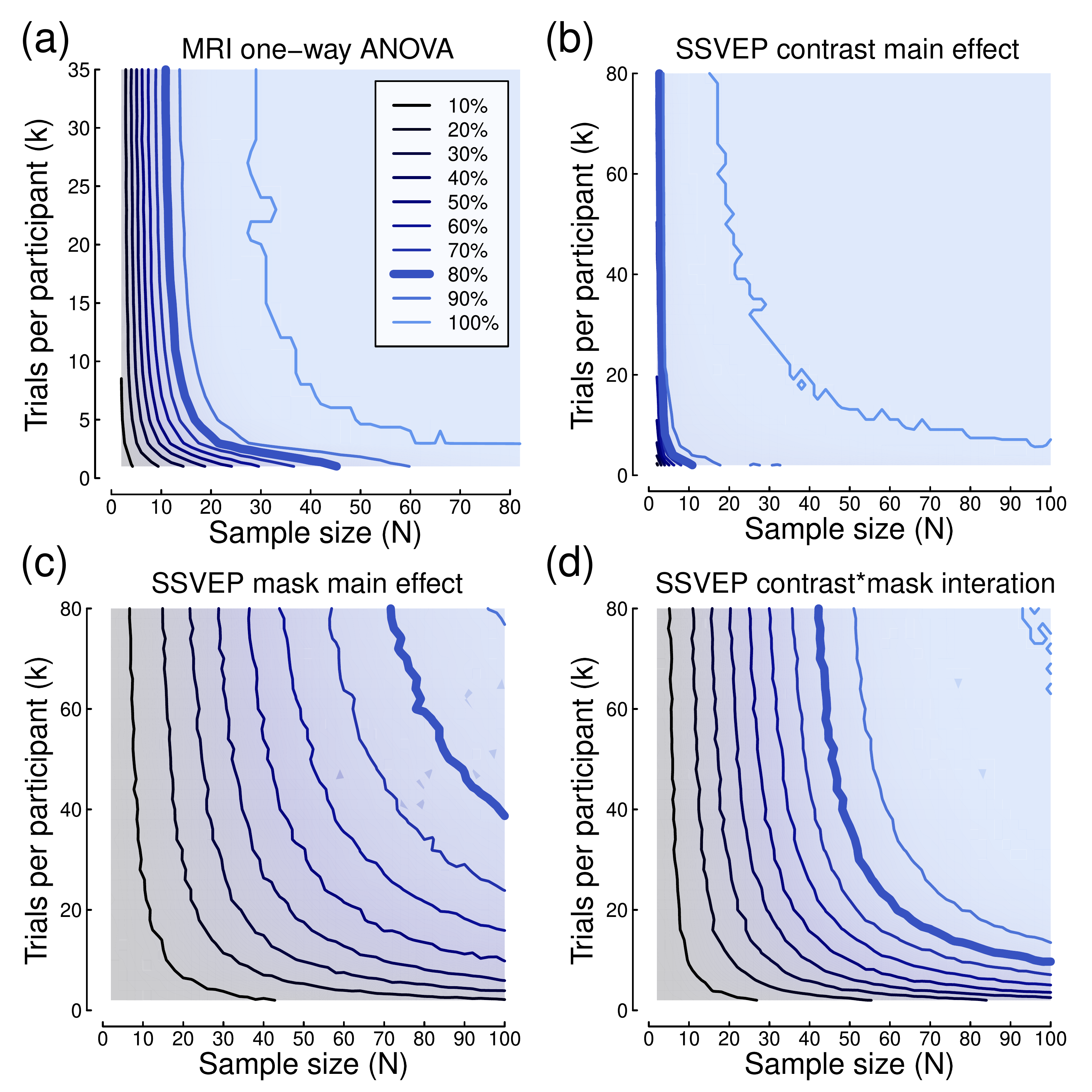}
\caption{Example power contours for one-way and factorial ANOVAs. Panel (a) shows a power contour plot for a one-way repeated measures ANOVA using three levels from the blocked fMRI data (summarised in Figure \ref{fig8}). Panels (b-d) show power contours for the main effects of contrast (b) and mask level (c), as well as their interaction (d) in a 7x2 repeated measures ANOVA design using the SSVEP data set (summarised in Figure \ref{fig6}).}
\label{fig12} 
\end{center}
\end{figure*}

In Figure \ref{fig12} we show power contour plots for repeated measures ANOVAs using two of the example data sets from the body of the paper. We conducted a one-way repeated measures ANOVA across the latter three TR times of the blocked design MRI experiment (using all five TR times produced such a large effect that the power contour analysis was uninformative). With the full data set, this produced a substantial significant effect (F$_{(2, 164)}$ = 40.39, \emph{p} $<6\times10^{-15}$, equivalent \emph{d} = 1.4). We then subsampled the data 10,000 times, repeating the ANOVA on each subsampled data set and calculating the proportion of significant tests (i.e. the power) to generate power contours. Figure \ref{fig12}a shows the power contour plot generated from this analysis, which closely resembles the power contour plots calculated for paired comparisons between these three conditions (Figure \ref{fig8}i,j).

We next conducted a factorial repeated measures ANOVA on data from the SSVEP experiment. As shown in Figure \ref{fig6}b, the experiment involved seven stimulus contrast levels. Participants also repeated all contrast conditions with an added orthogonal mask at high contrast. The two factors were therefore stimulus contrast (0 - 64\%), which produced a highly significant effect (F$_{(6, 1287)}$ = 171.83, \emph{p} $<2\times10^{-16}$, equivalent \emph{d} = 1.78) and mask contrast (0 and 32\%) which produced a smaller effect (F$_{(1, 1287)}$ = 12.89, \emph{p} $<0.0004$, equivalent \emph{d} = 0.19). The interaction between the two factors was also signficant (F$_{(6, 1287)}$ = 7.74, \emph{p} $<4\times10^{-8}$, equivalent \emph{d} = 0.35). Power contours for both main effects and the interaction are shown in Figure \ref{fig12}b-d. The main effect of stimulus contrast was so substantial that high power could be achieved with almost any combination of sample size and number of trials. The main effect of mask and the interaction were weaker, and again show the familiar tradeoff between N and \emph{k}. In practical settings, one should design an experiment to detect the smallest effect of interest with the desired power. For this example, the main effect of mask has the smallest effect, and so a replication of this experiment could use values along the 80\% contour in Figure \ref{fig12}c: for example, 100 participants each completing 40 trials, or 75 participants each completing 80 trials. Alternatively, if only the interaction were of theoretical interest, one could base the design on the constraints shown in Figure \ref{fig12}d.

For time-varying data using EEG and MEG (see Figures \ref{fig5} \& \ref{fig9}), it is commonplace to use cluster correction algorithms to control for multiple comparisons \citep[e.g.][]{Maris2007}. Informative power contours could in principle be constructed for significant clusters using either the number of trials (as here), or the number of time-points included within a cluster. Similar approaches might be applied to fMRI data, where the number of voxels included in a spatial cluster or a region of interest (ROI) will likely affect statistical power.

One limitation of the methods presented here is that they assume that trials are random, and independent of each other. In many paradigms, participants might become better at a task with practise (for example they could become more accurate, or their reaction times could speed up), or become fatigued after long testing sessions. This will place limits on the improvements gained by running additional trials, however the likely impact will vary across paradigms (see Figure \ref{fig3}c for an example). For large data sets it may be possible to estimate the nonstationarity of $\sigma_{w}$, and the impact this has on power \citep[see e.g.][]{Oertzen2013}. Other work has modelled multiple sources of variance in MRI studies explicitly using intra-class correlations \citep{Brandmaier2018}. This method permits dissociation of within-participant variance from various sources of measurement noise such as differences in variance between time of day, scanner model, and so on. Accurate estimates of relevant sources of variance will improve the overall accuracy of power analysis, which is particularly important given recent meta-analytic evidence \citep{Elliott2019} that test-retest reliability for task-based fMRI is typically very low (mean intra-class correlation < 0.4). 

An alternative to null hypothesis significance testing is the Bayesian approach. Bayesian alternatives to t-tests often calculate a Bayes Factor \citep{Rouder2009} as a test statistic, which indicates the relative probabilities of obtaining the observed data given the experimental and null hypotheses. For a given experimental design, one could calculate `Bayes factor contours' in an analogous manner to power contours, to estimate the number of trials and participants necessary to reach a specified level of evidence in support of one or other hypothesis. As Bayesian methods become more widespread, this may prove a useful alternative to traditional power analysis.

Another Bayesian-inspired method is to adaptively deploy data collection in the direction required to supply useful evidence to inform the outcome (posterior). An early example is the \emph{Quest} algorithm \citep{Watson1983}, used widely in psychophysics, which chooses the optimal stimulus level on each trial to provide the most information about the location of the threshold. Related methods have also been used to optimize data collection in fMRI experiments \citep{Lorenz2017}. Typically such approaches operate at a per-participant level, and will result in efficient use of the time available. If the ultimate aim is to combine results statistically across participants, then power contours might still be used to optimise the number of trials, in a similar fashion to that shown here for the contrast detection data (Figure \ref{fig4}), which also involved an adaptive (staircase) procedure. On the other hand, if the algorithm is designed to continue until particular conditions are met, traditional power analysis based only on sample size may be more appropriate.

Most discussion of power analysis is focussed on studies which involve statistically demonstrating the presence of some effect. However an alternative approach common in perceptual and cognitive research is to explain and predict patterns of response across multiple conditions using a computational model. In this tradition, each participant can be considered an independent `replication' of the phenomena under study \citep[see e.g.][]{Smith2018}, and the emphasis is on improving data quality through conducting many trials for each participant. Power contours might not be especially helpful under such circumstances, though knowledge of the within-participant standard deviation will inform decisions about how many trials to conduct.

Whereas experimental studies of the type we discuss here typically aim to reduce the sample variance ($\sigma_{s}$) in order to increase effect size, studies using individual differences approaches aim to maximise meaningful variation between participants. However, it is important that the observed variation ($\sigma_{s}$) is truly a result of individual differences (high $\sigma_{b}$) and not merely a consequence of poor measurement (high $\sigma_{w}$ and low $k$). Traditional psychometric instruments, such as tests of personality and ability, typically have high test-retest reliability, which implies low within-participant variance ($\sigma_{w}$), yet this may not be so for neuroscience and experimental psychology paradigms \citep[e.g.][]{Zuo2019,Elliott2019}. Estimating these values explicitly (e.g. using equation \ref{eq2}) may help individual differences researchers using such methods to optimise the number of trials and sample size to this end. We note that since $\sigma_{w}>\sigma_{b}$ for all estimates of these two parameters in the paradigms considered here (Table \ref{table:effects} and Figure \ref{fig10}b), individual differences studies will require sufficient trials to reduce the unwanted influence of intra-individual variability ($\sigma_{w}$) on sample variance ($\sigma_{s}$).

\subsection{Conclusions}

Here we present the rationale for incorporating the number of measurements (trials) into calculations of statistical power in experimental studies of psychology and human neuroscience. Power contour plots can be generated by subsampling existing data sets or using an online tool, and permit researchers to make informed choices about how many participants to test, and how long to test each one for, at the study design stage. However, as with all \emph{a priori} power calculations, the true effect sizes and variances will remain speculative until data have been collected.

\section{Acknowledgements}
We are grateful to everyone involved in collection of the data sets reanalysed here, and particularly to those who made their data publicly available. This work was supported in part by a Wellcome Trust (ref: 105624) grant, through the Centre for Chronic Diseases and Disorders (C2D2) at the University of York, awarded to DHB. Data collection and sharing for part of this project was provided by the Cambridge Centre for Ageing and Neuroscience (CamCAN). CamCAN funding was provided by the UK Biotechnology and Biological Sciences Research Council (grant number BB/H008217/1), together with support from the UK Medical Research Council and University of Cambridge, UK. We also thank Tom Hartley for helpful comments and for suggesting inclusion of the Iowa Gambling Task data set, and all those who offered constructive suggestions based on the preprint.

\bibliography{PCreferences.bib}

Data and scripts are available at: http://dx.doi.org/10.17605/OSF.IO/EBHNK

\end{document}